\documentclass{article}
\usepackage[utf8]{inputenc}
\usepackage{amsmath}

\def\code#1{\textsc{#1}}

\usepackage{relsize}
\usepackage{graphicx}
\usepackage{url}

\def\code#1{\texttt{#1}}
\usepackage{color}
\definecolor{dkgreen}{rgb}{0,0.6,0}
\definecolor{gray}{rgb}{0.5,0.5,0.5}
\definecolor{mauve}{rgb}{0.58,0,0.82}

\definecolor{dkgreen}{rgb}{0,0.6,0}
\definecolor{dkblue}{rgb}{0,0,0.6}
\definecolor{gray}{rgb}{0.5,0.5,0.5}
\definecolor{mauve}{rgb}{0.58,0,0.82}

\usepackage{xcolor}
\definecolor{commentgreen}{RGB}{2,112,10}
\definecolor{eminence}{RGB}{108,48,130}
\definecolor{weborange}{RGB}{255,165,0}
\definecolor{frenchplum}{RGB}{129,20,83}

\usepackage{listings}
\newtheorem{observation}{Observation}[section]

\title{GPU-friendly, Parallel, and (Almost-)In-Place Construction of
  Left-Balanced k-d Trees}

\author{Ingo Wald}
\date{Revision 3 (ArXiv v4) -- April 4, 2023}

\begin{document}

\maketitle
  \vspace{-2em}
\begin{abstract}
\noindent We present an algorithm that allows for building left-balanced and
  complete k-d trees over k-dimensional points in a trivially parallel
  and GPU friendly way. Our algorithm requires exactly one int per
  data point as temporary storage, and uses $O(\log N)$ iterations,
  each of which performs one parallel sort, and one trivially parallel
  CUDA per-node update kernel.
\end{abstract}

\section{Introduction}


K-d trees are powerful, versatile, and widely used data structures for
storing, managing, and performing queries on k-dimensional data.  One
particularly interesting type of k-d trees are those that are
left-balanced and complete: for those, storing the tree's nodes in
in-order means that the entire tree topology---i.e., which are the
parent, left, or right child of a given node---can be deduced from
each node's array index. This means that such trees require zero
overhead for storing pointers etc, which makes them particularly
useful for large data, or devices where memory is precious (such as
GPUs, FPGAs, etc).

This paper is about building such trees, in a efficient,
parallel, and GPU friendly way. Throughout the rest of this paper we
will omit the terms "left balanced" and "complete", and simply refer
to "k-d trees"; but always mean those that are both left-balanced and
complete.

\begin{figure}[ht]
  \vspace{-1em}
  \centering
  \begin{tabular}{cc}
    \begin{minipage}{.48\textwidth}
      \vspace*{-5cm}
      \includegraphics[width=.99\textwidth]{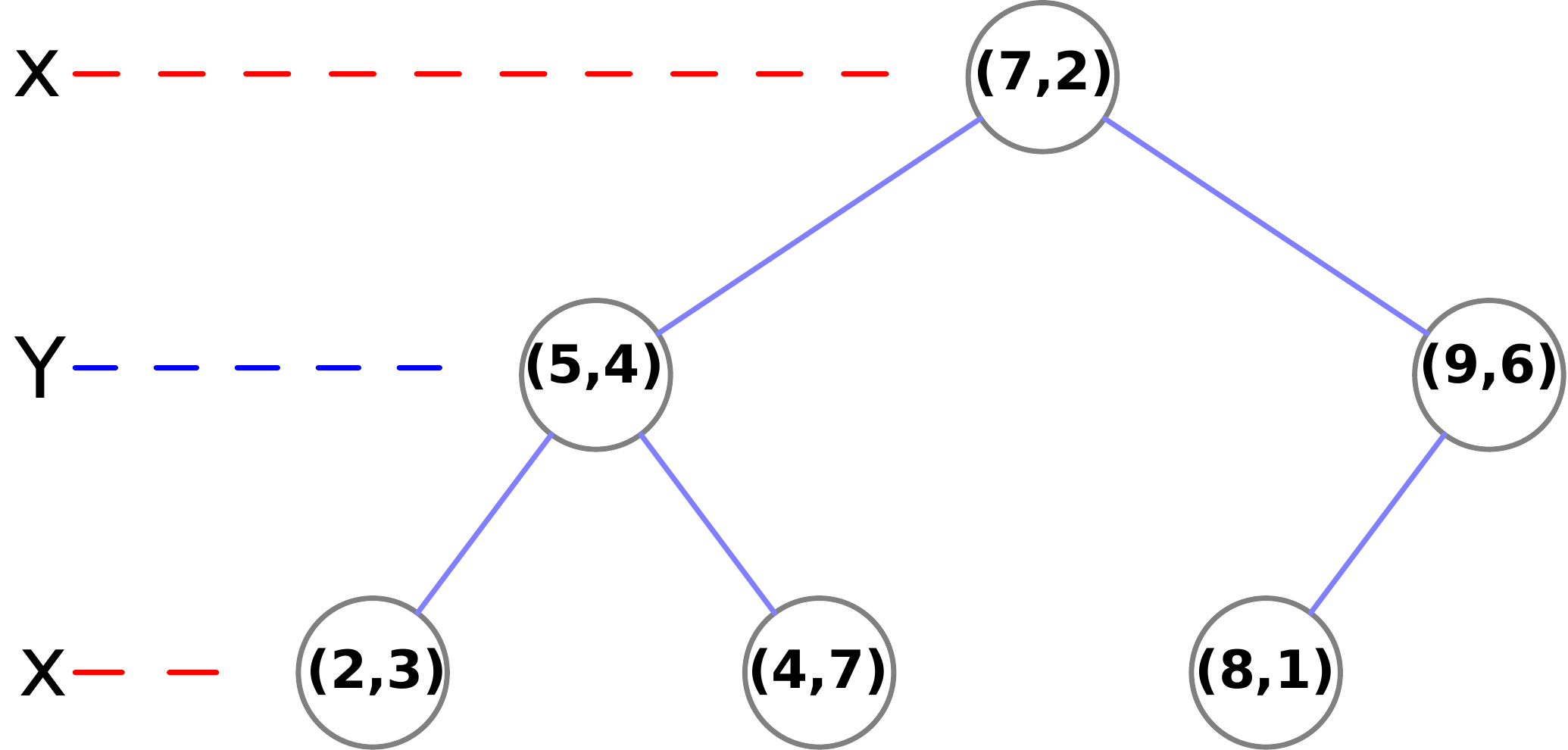}
    \end{minipage}
    &
    \includegraphics[width=.4\textwidth]{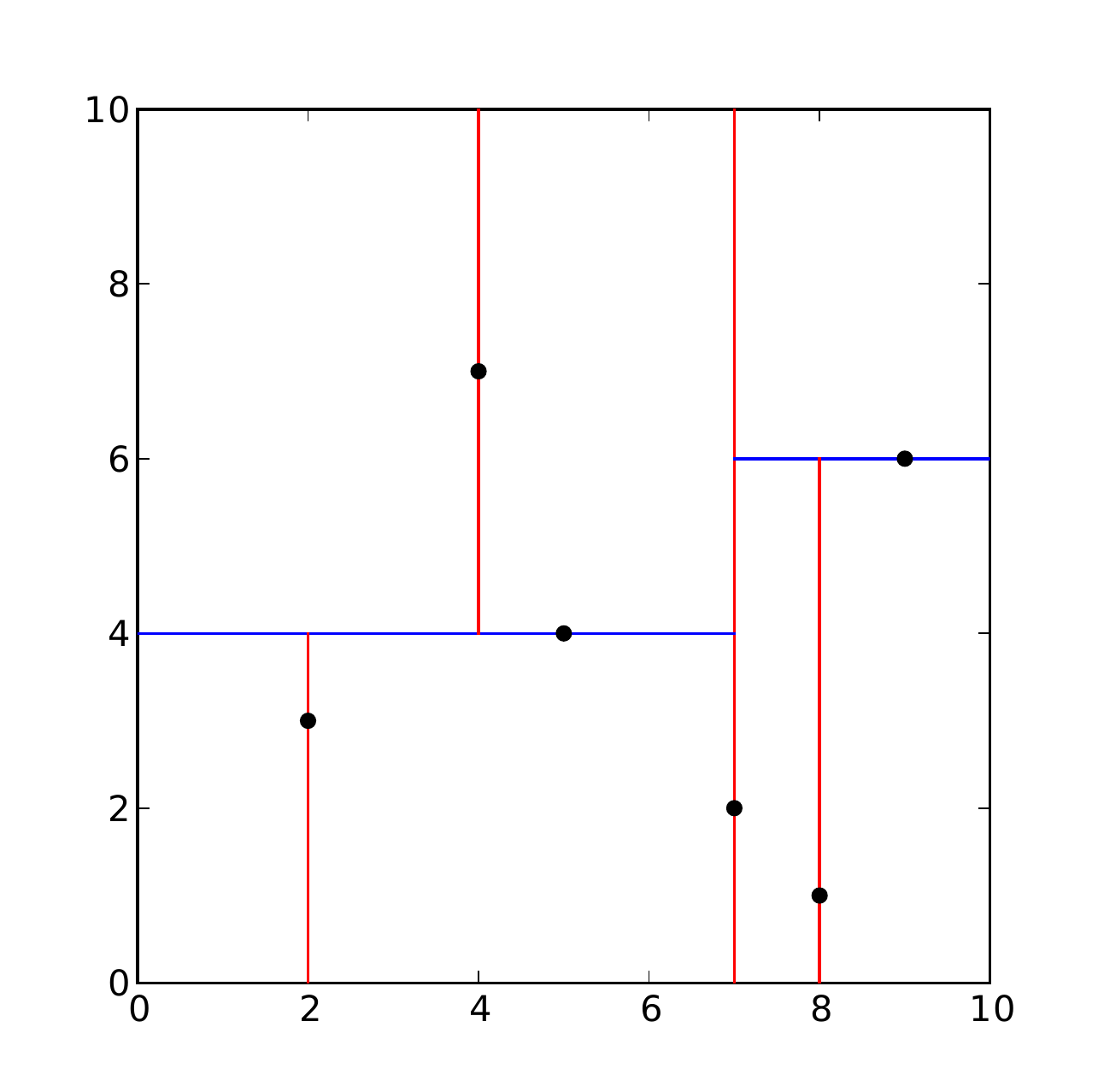}
  \end{tabular}
  \vspace{-2em}
  \caption{\label{fig:kd} Illustration of a 2-dimensional k-d tree
    from Wikipedia~\cite{wiki-kdtree}.
    Left: The balanced tree for
    2D point
    set $\{(2,3), (5,4), (9,6), (4,7), (8,1), (7,2)\}$.
    Right: The space partitioning achieved by that tree. }
\end{figure}

The problem with building k-d trees is that any node within this tree
has to simultaneously fulfill two different conditions: first, the k-d
tree condition that all nodes in its left (respectively right)
sub-tree are, relative to that node's partition plane, all $\leq$
respectively $\geq$ that node's coordinate (in that node's chosen
dimension); and second, the balancing condition that the resulting
tree must be left-balanced (which implicitly fixes how many nodes any
given sub-tree has to have). Building a k-d tree over a given set
of $N$ elements thus requires finding an ordering of these elements
such that every node fulfills both of these conditions; and the problem
of finding this ordering becomes particularly interesting when dealing
with highly parallel architectures such as GPUs.

The usual way to build such trees is to start with a single array of
nodes, then sort this array, split it at the appropriate position that
ensures balancing, and recurse; but this approach has several issues:
one is that the sorting steps require array order, but the tree
requires level order, typically requiring to keep two copies of the
data. An even worse problem is that running this algorithm in parallel
requires significant effort in tracking which sub-trees still need to
be built, with which plane, with which elements, etc.  Finally, like
many other top-down hierarchical build algorithms this approach
suffers from the fact that the granularity of parallelism is changing
radically over time, from $O(1)$ jobs with $O(N)$ elements at first,
to $O(N)$ jobs of $O(1)$ elements at the end; this causes issues
because the ``same'' logical operation on a sub-tree (such as, for
example, sorting its elements) may require very different
implementations across different stages of the algorithm (e.g.,
sorting a list of 4 elements requires a different algorithm than
sorting one with 4 million). This need to specialize for problem
size often leads to complex code, lots of special cases, etc.

In this paper, we describe an algorithm that can transform a set of
$N$ k-dimensional data points into a proper left-balanced complete k-d
tree by performing a series of $\log_2{N}$ steps of re-arranging these
points.  This algorithm can best be visualized as starting each node
in the root of the tree, and having them "trickle" down the tree until
each node has reached the location in the tree where both k-d tree and
balancing properties are fulfilled. In particular, our algorithm can
do that without having to require a secondary copy of the data set,
without having to create, partition, or manage temporary lists of
nodes during construction, and in a way where every kernel and sort
operation always operates on the entire set of all data points, in a
way that trivially lends to parallel execution.

\section{Definitions and Preliminaries}

Throughout this paper we assume we need to build a left-balanced binary
k-d tree over $N$ elements of $k$-dimension data points.  Though we will
later remove this restriction we will also, for now, assume that the
k-d tree to be built will, in any level $l$, partition a sub-tree's
data points using the dimension $l\bmod k$.

When referring to nodes in the k-d tree, we use the canonical way of
doing so via that node's array index if the tree is stored in
level-order. I.e., we refer to the root node as node 0, its children
as nodes 1 and 2, etc.\  Using that indexing, many important properties
for a given node can be computed simply from a given node's index. For
example, the parent and children of node $i$ are
$\code{parent}(i)=(i-1)/2$, $\code{lChild}(i)=2i+1$, and
$\code{rChild}(i)=2i+2$, etc.\

In addition to parent-child relationships we can also compute
information about the tree in general. For example, we can compute the
tree level of a node $i$ as
$\code{level}(i)=(\code{int})log_2(i+1)$---or in integer bit-arithmetic,
as $\code{level}(i)=31-\code{clz}(i+1)$, where \code{clz} is the instruction
for counting the number of leading zeroes in a 32-bit int type (which
every modern CPU or GPU supports). We can similarly compute the total
number of levels $L$ in a tree of $N$ nodes as
$L=\code{numLevels}(N)=\code{level}(N-1)+1$; or the number of nodes in a full tree of $l$
levels as $F(l)=2^l-1$ (or $F(l)=(1\ll l)-1$ using integer bit-arithmetic);
etc. Finally, in addition to nodes we can also argue about
\emph{sub-trees} within a given tree, which we refer to through the
index of the node at the root of that sub-tree; i.e., the sub-tree
$S(i)$ is the set of nodes that are within the sub-tree rooted at node
$i$. We will derive some important properties for sub-trees below.


\subsection{Sub-tree-Size and Pivot Element}

One particularly interesting value we will be making use of in our
algorithm is the value of the \emph{number of nodes in a sub-tree}
under a given node \code{s}, which we will refer to as $\code{ss}(s)$
(for \emph{subtree-size} of sub-tree $s$). We show below that for any
left-balanced tree of $N$ nodes this value can be computed, for any
$s$, in $O(1)$ complexity (Section~\ref{sec:ss}).

The reason this value is important is that for any given sub-tree $s$
we can use this to find the data elements within this sub-tree:
assuming we have a sequential list of those elements, and assuming
these are already sorted based on the proper split dimension desired
for node $s$, then by virtue of the k-d tree and balancing
requirements we know that the first $\code{ss}(\code{lChild}(s))$
nodes must be in the left sub-tree, the next one must be the root
(i.e., it must be the element that in the eventual level-order layout
of the correct final k-d tree should be at array-position $s$), and
all others must belong to the right sub-tree.  This position---the one
that splits this list into left sub-tree nodes, root, and right
sub-tree nodes---we will also refer to as the pivot position, and the
element at that location as the pivot element.


\begin{observation}
\label{observation-pivot}
Let $S(s)=S_0,S_1...$ be the sequential list of data points known to
form the set of points that make up the sub-tree under $s$, and let
these points be sorted by their $(\code{level}(s) \bmod k)$'th
coordinate. Then, the value $\code{ss}(\code{lChild}(s))$ is the k-d
tree pivot position for this list; the element at that position should
be the node $s$ of the final tree, those with lower indices are those
that belong into the left sub-tree under $s$, and those with higher
indices belong into the right sub-tree.
\end{observation}

\subsection{Level-l Ancestors of Nodes and Data Points}

A second important value we will make use of in our algorithm is the
\emph{level-l ancestor} of a node $n$ (or $\code{lla}(n)$.  For any
given k-d tree node $n$ with $\code{level}(n)\leq l$ the level-$l$
ancestor is the index of the (uniquely defined) node $n'$ that lives
on level $l$ of the tree, and whose sub-tree $n$ is in. For values $l
> \code{level}(n)$, we define the level-$l$ ancestor of $n$ to be
$n$ itself. I.e., the level-$0$ ancestor of any node is the
root node 0; the level-1 ancestor of any node $n$ is either 1 or 2
(except for $n=0$, whose \code{lla} is $0$), etc.

Just like we can argue about level-$l$ ancestors of a given
\emph{existing} k-d tree of $N$ elements, so we can also argue about
level-$l$ ancestors of \emph{data points} that may or may not yet be
in the proper k-d tree order: in this case, we use the term level-$l$
ancestor of a given input data point $p$ to mean the \code{lla} of the
node $n$ that $p$ \emph{should} be stored in if that array of data
points were already in a valid balanced k-d tree order.  The reason
this value is important is that by definition, if we know the
\code{lla} for any data point $p$ to be $n_p$, and the level of this
$n_p$ is $\code{level}(n_p) \leq l$, then $n_p$ must be the position
that $p$ should be at to fulfill the k-d tree criteria. In particular,
if we know the level-$L$ ancestor of every data point $p$ (where $L$
is the total depth of the tree) then we know, for each data point,
exactly the position it should be at to form the proper k-d
tree.

\section{Algorithm}

The core idea of our algorithm is to look at the level-l ancestor
property of every input data point $p$, for values of $l$ that
increase by one in each iteration (and thus, effectively \emph{fixes}
one more level of the tree in each iteration). To do this we
temporarily associate each data point with a 32-bit integer value
(that we also call the \emph{tag} for that data point), and use that
to store each iteration's \code{lla} value for this node. We start by initializing each data
point's tag to 0 (which is, by definition, the level-$0$ ancestor of
every node), then in each iteration $l$ we read the node's current \code{lla}
value from its tag, then compute its respective \code{lla} value for
the next level $l+1$. Since the \code{lla} for levels greater than the
one that the node is on will not change any more, this effectively
fixes one more level of the tree in each iteration, meaning that after
$L=\code{numLevels}(N)$ steps each node's tag will store exactly the node ID that this
data point should be in.

The key to this algorithm is, obviously, the kernel that refines a
node's \code{lla} from one level $l$ to the next level $l+1$. To do
this, we rely on a suitably chosen ordering of all (node,tag)
pairs of data, that we will derive next. Using this ordering we
will---after putting all data points into that chosen order using a
parallel sort---then easily be able to derive each data point's
next-level \code{lla}.

\subsection{Sort Order and Properties of that Order}

In any given one of the $L$ iterations of our algorithm, let us
consider a comparison operator that consists of two nested orderings:
first, a \emph{major} sort order that sorts data points by ascending
tag value; and second, a \emph{minor} sort order that sorts those
elements \emph{with same tag} based on their respective $(l \bmod
k)$'th coordinate, (where $l$ is the iteration we are currently
performing (and in which we are, consequently, finding the 
position for all nodes on the $l$'th level of the tree).

In pseudo-code, we can use the following \code{less()}-operator to achieve
this chosen $l$'th step ordering of our data points:
\begin{lstlisting}
bool less(int idx_a, int idx_b, int l) 
{
  int dim = l % k;
  return
    (tags[idx_a] < tags[idx_b])
    || (tags[idx_a] == tags[idx_b])
    && (points[idx_a][dim] < points[idx_b][dim];
}
\end{lstlisting}


Let us now assume that in each iteration $l$ the tag for any given
data point $p$ will initially specify that point's \code{lla} for
level $l$. In that case, based on how that \code{lla} is defined we can make the following
\begin{observation}
\label{observation-numllas}
Let $T=t_0,t_1,....$ be an array of ints where each $t_i$ stores the
level-$l$ ancestor of a given node in a binary tree of $N$ nodes and
$L(N)>l$. By definition of the \code{lla}, there must be exactly
$F(l)$ tags that each point to a node $n'$ on a level $l'<l$; and all
other tags must point to a node on level $l$.
\end{observation}

If we further apply a sort using the above comparison operator, then
we immediately arrive at 
\begin{observation}
\label{observation-order}
Let $A=A_0,A_1...$ be the array of data points after the $l$'th
iteration's sorting step.  Then, the first $F(l)$ elements of the
array will be the nodes with tags $0,1,\dots,F(l)-1$, and those
will all be already at their proper locations in the array. Following
those $F(l)$ nodes for levels $[0..l)$, starting at offset $F(l)$, there will
be the $\code{ss}(F(l))$ nodes that will end up in sub-tree $F(l)$,
sorted by their $(l \bmod k)$'th coordinate. Following those, at offset
$F(l)+\code{ss}(F(l))$ there will be the $\code{ss}(F(l)+1)$ nodes with \code{lla}
$F(l)+1$ (also sorted in the proper dimension), etc.
\end{observation}

We can now use this observation to define an efficient \emph{tag update kernel}
that computes, from this array layout, each node's \emph{next-}level ancestor.

\subsection{Tag Update Kernel}

Using observation~\ref{observation-order} we know that for each
sub-tree $s$ on level $l$ we will find all nodes tagged with this
sub-tree in a single coherent region of our array; we call this the
\emph{segment} of the $s$-tagged array elements. Let us now call the
index where this segment with $s$-valued tags starts the \emph{segment
  begin} index for $s$, \code{sb(s,l)}. As this segment contains all
of the data points in the sub-tree under $s$, its size will obviously
be \code{ss(s)}.

Within each such segment, we can now use observation~\ref{observation-pivot} to
determine, for each element of this array, whether it is the pivot
element, or in the left respectively right sub-tree of $s$. Using this,
we arrive at a node update kernel
that--assuming the data is properly sorted as above--for
each array item $i$ in parallel, performs the following steps:
\begin{itemize}
\item check if $i < F(l)$. If so, the tag is already correct; terminate.
\item read the tag $s=s(i)$ from the tag array to know which sub-tree the data point is in.
\item compute $\code{sb}(s,l)$ and $\code{ss}(\code{lChild}(s))$ and use these to
  compute the pivot position $p(s)=\code{sb}(s,l)+\code{ss}(\code{lChild}(s))$.
\item for indices less than pivot position, change tag to $\code{lChild}(s)$;
  for those greater than change it to $\code{rChild}(s)$.
\end{itemize}

\medskip\noindent
This leads to the following pseudo-code:
\begin{lstlisting}
__global__ void updateTags(int tags[], Point points[], int N, int l)
{
   int arrayIdx = CUDA thread index;
   if (arrayIdx >= N || arrayIdx < F(l)) 
      /* invalid index, or already done */
      return;
   int currentTag = tags[arrayIdx]; // must be a node index on level l
   int pivotPos = sb(currentTag) + ss(lChild(currentTag))
   if (arrayIdx < pivotPos)
      tags[arrayIdx] = lChild(currentTag)
   else if (arrayIdx > pivotPos)
      tags[arrayIdx] = rChild(currentTag)
   else
      tag remains unchanged; this is the root of this sub-tree
}
\end{lstlisting}

\medskip\noindent
Using this tag update kernel, our whole algorithm is simply the following
set of $\log_2 N$ iterations:
\begin{lstlisting}
void parallelKDTreeBuild(Point points[N])
{
   int tags[N];
   in parallel: initialize all tags to 0;
   for (int l = 0 .. log2(N)) {
      parallel sort {tags,points}, using chosen less() w/ dim l%k
      in parallel: updateTags(tags,nodes,l)
   }
}
\end{lstlisting}

\medskip\noindent After $L=\code{numLevels}(N)$ steps, every one of
the data points will have a tag that indicates where in the array that
data point should be stored; and since the array will be sorted by
this tag every data point will automatically be at the proper
position---meaning that the resulting array is indeed the proper k-d
tree for these data points.





\section{Computing $\code{sb}(s)$ and $\code{ss}(s)$ in O(1)}

Throughout our algorithm there were two properties that we made
frequent use of: the \emph{segment begin} function $\code{sb}(s,l)$,
and the \emph{sub-tree size} function $\code{ss}(s)$. We have claimed
that each one of those can be computed with $O(1)$ complexity, but have not
yet described how to do that.

\subsection{Computing Size of Sub-tree $s$, \code{ss}(s)}
\label{sec:ss}

Let $\code{ss}(s)$ be the number of nodes in the sub-tree rooted in
node $s$. We observe that this values is an intrinsic property of any
k-d tree of $N$ nodes, not specific to our algorithm.  We also observe
that this is useful even for other k-d tree construction algorithms,
as it allows for computing the pivot-position even for list-based
recursive partitioning schemes.

Computing this value recursively is trivial, but costs $O(N)$. To
compute this in $O(1)$ we first observe that the tree is left-balanced
and complete; i.e., if $L$ is the number of levels in the tree then all
levels $l'$ with $0\leq l' \leq (L-2)$ are entirely full, and level $L-1$ is filled from the
left (i.e., it contains nodes $[2^{L-1}-1,...,N)$).

  Now let $l$ be the level of node $s$; then we know that all levels
  from $l$ to including $L-2$ must be entirely full; and those must
  contain $F((L-1)-l)$ nodes. Furthermore, we can also compute the range
  of nodes that this sub-tree \emph{would} have on level $L-1$, and
  compare that with the range of nodes that the tree actually has.

  The \emph{first} node of sub-tree $s$ on level $(L-1)$---which we
  call the \emph{first lowest-level child} (\code{fllc}(s))---can be
  computed by repeatedly computing \code{lChild} for a total of
  $((L-1)-l)$ times. Since $\code{lChild}(i)=2i+1$, calling this
  function once is synonymous with shifting $i$ by 1 to the left, and
  setting the last bit to $1$; i.e., we take the binary code of $i$
  and append a ``1'' bit. Doing this $L-l-1$ times on $s$ is then
  synonymous with taking the binary code of $s$ and appending $L-l-1$
  ``1'' bits, which in integer arithmetic can also be computed as
  \begin{lstlisting}
    fllc_s = ~((~s) << (L-l-1)).
  \end{lstlisting}

  \medskip\noindent
  Obviously, if this value is $\geq N$, then sub-tree $s$ has no nodes
  on level $L-1$; otherwise, it will have at most $N-\code{fllc\_s}$
  nodes. We also know that in a full tree the sub-tree under $s$ would
  only be $2^{L-l-1}$ nodes wide, so the total number of sub-tree-$s$
  nodes on level $L-1$ can be computed as
  \begin{lstlisting}
    numNodes_s_on_lowest_level = min(max(0,N-fllc_s), 1<<(L-l-1)).
  \end{lstlisting}

  \medskip\noindent
  Together with the full inner nodes, this for $\code{ss}(s)$ evaluates
  to
  \begin{lstlisting}
    fllc_s = ~((~s) << (L-l-1)).
    ss_s
    = /* inner nodes  */ 1<<(L-l-1) - 1
    + /* lowest level */ min(max(0,N-fllc_s), 1<<(L-l-1)).
  \end{lstlisting}

  \medskip\noindent
  In this expression, $l$ and $L$ can be computed in constant time using simple
  \code{clz} as described above; all other operations are 
  basic arithmetic operations like bit-shift, addition, and integer
  min/max, all of which are constant-cost and trivially cheap
  on modern architectures.

\subsection{Computing Segment-$s$-begin in step $l$, $\code{sb}(s,l)$}
\label{sec:sb}

The second important property our algorithm frequently relies on is
what we called the \emph{sub-tree-s-begin} index; i.e., the index in
our sorted array (in step $l=\code{level}(s)$) where all the data
points tagged to be in sub-tree $s$ can be found (we observe that in
any iteration $l$ our algorithm calls this function only on sub-trees
$s$ that are in the $l$'th level).

Recalling observation~\ref{observation-order} we know that in the
$l$'th step the array will first contain the $F(l)$ nodes of the top
$l$ levels that have already found their final array position; followed
by those for the first sub-tree in level $l$ (tagged with
$F(l)$); then those for sub-tree $F(l)+1$; then those for sub-tree $F(l)+2$;
etc.  Since we can compute the size of any sub-tree $s$ through
$\code{ss}(s)$, we could compute $\code{sb}(s,l)$ through
$$\code{sb}(s,l) = F(l)+\sum_{i=F(l)}^{s-1}\code{ss}(i),$$
but this would have $O(N)$ complexity.

To compute that same value in $O(1)$ we refer to the same observations
we made in the previous section. First, we observe that there are a
total of $\code{nls}=s-F(l)$ sub-trees on level $l$ that are to the
left of $s$ (and thus come before $s$); we call this the \emph{number
  of left siblings of $s$}, \code{nls(s)}.
Each of these sub-trees must have $L-l-1$
levels that are guaranteed to be full, and as there are
$\code{nls}(s)$ of those there must be exactly $\code{nls}(s)F(L-l-1)$
such nodes on those inner levels, across all those trees to the left of $s$.  On the
lowest level of the tree, these left siblings fill the tree from the
left, so the number of nodes on the lowest level must be the lesser 
of how many nodes these trees \emph{would} have if they were all
filled (which is $\code{nls}(s)2^{L-l-1}$), and how many the actual
tree has on that level (which is $N-F(L-1)$).

Taken together, this combines to
\begin{lstlisting}
  /* typically have those already, so just for completeness: */
  /* level of s */ l = 31-clz(s+1)
  /* num levels */ L = 32-clz(N)

  nls_s = s - ((1<<l)-1)
  sb_s_l
  = /* top l levels */ (1<<l)-1
  + /* left siblings, inner  */ nls_s*(1<<(L-l-1)-1)
  + /* left siblings, lowest */ min(nls_s*(1<<(L-l-1)),
                                    N-((1<<(L-1))-1))
\end{lstlisting}
Like for $\code{ss}(s)$, this can be computed with just a few
\code{clz}s, shifts, and basic arithmetic operations, in $O(1)$.

\section{An Example Algorithm-Walkthrough}

To illustrate how this algorithm performs in practice, we have run our
reference implementation of it on 10 randomly chosen \code{int2} data points, and use this
section to show--step by step--how it modifies the array of data
points and tags.

In our example, let us assume we start with the 10 randomly chosen
points
{\relsize{-1}{$((10,15),(46,63),(68,21),(40,33),(25,54),(15,43),(44,58),(45,40),(62,69),(53,67)$}}.
Let us attach each of these with a tag, initialize each such tag with
0, and show these in table form; this gives an initial state as
\begin{lstlisting}
   array-idx:    0    1    2    3    4    5    6    7    8    9
   tags     :    0    0    0    0    0    0    0    0    0    0
   data[i].x:   10   46   68   40   25   15   44   45   62   53
   data[i].y:   15   63   21   33   54   43   58   40   69   67
\end{lstlisting}

\par\noindent Let us now run iteration $0$, which---since all nodes have the same
tag of 0---will first result in all nodes being sorted in $x$:
\begin{lstlisting}
   // after step 0 sort
   array-idx:    0    1    2    3    4    5    6    7    8    9
   tags     :    0    0    0    0    0    0    0    0    0    0
   data[i].x:   10   15   25   40   44   45   46   53   62   68
   data[i].y:   15   43   54   33   58   40   63   67   69   21
\end{lstlisting}

\par\noindent Running the update-tags steps will evaluate the
pivot-pos for sub-tree $0$ (which in this stage all nodes will be in)
as $\code{ss}(\code{lChild}(0))$, which is $6$, leading to these
updated tags
\begin{lstlisting}
   // after step 0 tags updated
   array-idx:    0    1    2    3    4    5    6    7    8    9
   tags     :    1    1    1    1    1    1    0    2    2    2
   data[i].x:   10   15   25   40   44   45   46   53   62   68
   data[i].y:   15   43   54   33   58   40   63   67   69   21
\end{lstlisting}

\par\noindent For $l=1$ sorting now already ``pins'' element $0$, and
sorts elements by sub-tree ID $0$ and $1$, respectively, and with
ascending $y$ values within the same sub-tree:
\begin{lstlisting}
   // after step 1 sort
   array-idx:    0    1    2    3    4    5    6    7    8    9
   tags     :    0    1    1    1    1    1    1    2    2    2
   data[i].x:   46   10   40   45   15   25   44   68   53   62
   data[i].y:   63   15   33   40   43   54   58   21   67   69
\end{lstlisting}

\par\noindent Updating tags in step $l=1$ now finds nodes $1$ and $2$,
and updates their children into sub-trees 3--6, respectively:
\begin{lstlisting}
   // after step 1 tags updated
   array-idx:    0    1    2    3    4    5    6    7    8    9
   tags     :    0    3    3    3    1    4    4    5    2    6
   data[i].x:   46   10   40   45   15   25   44   68   53   62
   data[i].y:   63   15   33   40   43   54   58   21   67   69
\end{lstlisting}

\par\noindent Sorting for iteration $l=2$ then yields nodes
0--2 fixed correctly, and sub-trees 3--6 in proper order:
\begin{lstlisting}
   // after step 2 sort
   array-idx:    0    1    2    3    4    5    6    7    8    9
   tags     :    0    1    2    3    3    3    4    4    5    6
   data[i].x:   46   15   53   10   40   45   25   44   68   62
   data[i].y:   63   43   67   15   33   40   54   58   21   69
\end{lstlisting}

\par\noindent Updating tags now builds all tags for the third level:
\begin{lstlisting}
   // after step 2 tags updated
   array-idx:    0    1    2    3    4    5    6    7    8    9
   tags     :    0    1    2    7    3    8    9    4    5    6
   data[i].x:   46   15   53   10   40   45   25   44   68   62
   data[i].y:   63   43   67   15   33   40   54   58   21   69
\end{lstlisting}

\par\noindent
Finally, performing a final sort gives the (correct) final result as
\begin{lstlisting}
   // after final sort
   array-idx:    0    1    2    3    4    5    6    7    8    9
   tags     :    0    1    2    3    4    5    6    7    8    9
   data[i].x:   46   15   53   40   44   68   62   10   45   25
   data[i].y:   63   43   67   33   58   21   69   15   40   54
\end{lstlisting}

\par\noindent
The algorithm is now complete, and this is indeed the correct k-d tree
array for this input.

\section{Extension to Split-Widest-Dimension}

In our discussion above we have assumed that for each node $s$ the
builder would use dimension $l\bmod k$ for partitioning, where $l$ is
the level of the tree that $s$ is in. Though this is exactly what many
users of k-d trees want to do, for some applications it has been shown
that always partitioning each sub-tree along the dimension where its
associated region of space has widest extent can lead to significantly
better query performance.

To do this, a top-down recursive builder would first compute the
bounding box of the entire data set (which is the \emph{domain} of the
entire tree, and the region of space that the root node is going to
partition); then in each step the builder would consider the domain of
the current sub-tree, choose the dimension where this domain's bounding box is
widest, and use that for partitioning. It would then compute the
domains of the left and right sub-tree based on the parent's domain 
partitioning plane, and recurse.

In our algorithm we cannot easily  track bounding boxes for
each sub-tree in a top-down manner, as we have potentially many such sub-trees in flight at
any point in time. We can, however, easily compute a given sub-tree's
bounding box by following its path from the root \emph{backwards}:
doing so will also iterate over all the partitioning planes ``above''
s, and in each step, we know whether $s$ is to the left or or right of
this tree. Thus, for any $s$ we can start with the (pre-computed) bounding box of the
entire domain, then walk from $s$ upwards to the
root, and in each step ``clip'' the bounding box by the given
ancestor's splitting plane.

With this in mind, we modify our algorithm as follows: We first
reserve the lower $\log_2 k$ bits in the tag value to store the
partitioning index we want to use for the sub-tree $s$ that this node
is in. In the initialization step we pre-compute the bounding box of
the entire data sets, and initialize these bits to the widest
dimension of this data set bounding box.

In the sort step, we proceed as above, except that for any comparison
between points $A$ and $B$ we first retrieve $A$'s node tag, extract
that tag's stored split dimension, and, for similar tag compare $A$
and $B$ across this dimension. We observe that it can absolutely
happen that this comparison operator gets called with two nodes $A$
and $B$ that have different dimensions stored in their tag; however,
this is perfectly OK: the comparison operator first compares by tag,
so the comparison by coordinate only ever happens if these two nodes
have the same tag---in which case they are in the same sub-tree and use
the same split dimension.

The last modification we have to do is make sure that the tag update
kernel will properly compute the split dimension for each of the next
step's sub-trees. To do that we first observe that this has to be done
only for those nodes whose new tag $s'$ got updated to either
$s'=\code{lChild}(s)$ or $s'=\code{rChild}(s)$.  If that is the case,
we first compute the bounding box of sub-tree $s'$: we start with the
scene bounding box, and first clip that by the split plane of $s$,
which this step has just determined, and which thus is now known. We
then follow $s$ up the tree, and further clip this bounding box by the
visited ancestors (all of which have already been
fixed in previous iterations). The result of this is, for each $n$, the bounding
box of the newly sub-tree $s'$ that this node just got tagged with. We
then compute the proper split dimension, and store this in the new
tag.

This step of computing the split dimension for each node obviously
increases the cost of the tag update step from $O(1)$ to $O(\log N)$
per node, and therefore, to $O(N \log N)$ for the entire update step
across all nodes.  However, the cost of this inner loop was $O(N \log
N)$ even before this change (due to the cost for sorting), so the
total algorithm's complexity remains unchanged at $O(N \log^2
N)$. Sample code for this algorithm is provided together with sample
code for the round-robin based algorithm, at
\url{https://github.com/ingowald/cudaKDTree}.

\section{Discussion}

In this paper we have described an algorithm that allows for the
parallel, GPU-friendly, and (mostly-)in place construction of
left-balanced k-d trees. Sample code for both round-robin and
split-widest-dimension, as well as for different data types and
queries operating on those trees, can be found at
\url{https://github.com/ingowald/cudaKDTree}~\cite{github-repo}.

Our algorithm can build left-balanced k-d trees (of either round-robin
or split-largest-dimension variety) in $O(N log^2 N)$.  We call our
algorithm \emph{mostly-}in place because our algorithm itself only needs
a constant one int per node in temporary storage, independent of
number of nodes, number of dimension, size of other payload stored
with each data point, etc. Of course, we also rely on sorting, and
based on used sorting algorithm this may require additional temporary
storage; in particular the thrust~\cite{thrust} based sort we use in
our sample code \emph{will} use non-trivial amounts of storage. A sort
based solely on swapping---for example, \emph{bitonic
  sort}~\cite{bitonic-sort}---would avoid this (and quite possibly be
faster, too), but for the sake or readability has not been used in our
reference implementation.

\subsection{Performance}

Our algorithm is designed for parallel execution, and reasonably fast;
on a modern NVIDIA-3090TI GPU a k-d tree over 10 million points can be
built in well under one second (see Table~\ref{tab:perf}). That said,
we make no claims that our algorithm (and in particular, our
implementation) was necessarily the fastest way such trees can be
built. As for memory usage, the main cost factor in terms of compute
is parallel sorting: sorting on GPUs can be incredibly fast for radix
sort-style algorithms that operate on simple scalar types; but for
current libraries sorting with custom operators is much slower. This
could potentially be addressed by using a different implementation of
parallel sort, and/or by using a 64-bit tag in which the upper bits
store sub-tree index and lower bits replicate the point's coordinate
in the chosen sorting dimension (one can then use a key-value radix
sort with a 64-bit integer); but for didactic reasons this has not
been done in our reference implementation.

\begin{table}[h!]
  \centering
  \begin{tabular}{c|ccccccc}
    N          & 1K     & 10K   & 100K  &1M     & 10M   & 100M  & 1B\\
    \hline
    \hline
    \multicolumn{8}{c}{\code{int1} data type}\\
    \hline
    RTX~8000   & $<1$ms & 0.6ms & 1.4ms & 12.8ms & 165ms & 1.89s & 24.7s \\
    A-6000     & $<1$ms & 0.6ms & 1.2ms & 11.9ms & 124ms & 1.37s & 17.7s \\
    RTX~3090TI & $<1$ms & 0.6ms & 1.2ms & 11.0ms & 106ms & 1.07s & 16.8s \\
    \hline
    \hline
    \multicolumn{8}{c}{\code{float4} data type}\\
    \hline
    RTX~8000   & $<1ms$ & 0.9ms & 6.0ms & 54.4ms & 783ms & 10.3s & (oom) \\
    A-6000     & $<1ms$ & 0.9ms & 5.6ms & 49.7ms & 536ms &  6.5s & (oom) \\
    RTX~3090TI & $<1ms$ & 0.9ms & 5.6ms & 44.4ms & 424ms &  5.3s & (oom) \\
  \end{tabular}
  \caption{\label{tab:perf}Build performance for given number of
    \code{float4} data points ($K$ means $1,000$, $M$ means 1 million,
    and $B$ means 1 billion), using round-robin partitioning and
    uniformly distributed random points in $[0,1]^4$ (averaged over
    1,000 runs for $N<1M$, over 100 runs each for $1M \leq N < 1B$,
    and across 10 runs for $N=1B$). \code{oom} means ``out of memory''}
\end{table}

We also observe that there are some obvious savings in that in each
$l$'th step the sort and tree update kernels can be omitted for the
first $F(l)$ elements of the array; and in that one does not actually
need to build the lowest level of the tree since all sub-trees on that
level are of size 1, and after sorting are already at the right
position. These simple optimizations have already been done in our
implementation.

\subsection{Simplicity and Generality}

Arguably the biggest advantage of our algorithm is its simplicity:
though the computation of values like $\code{ss}(s)$ and
$\code{sb}(s,l)$ at first glance does not look simple at all, once
these are carefully derived (see above) the rest of the algorithm is a
trivially simple sequence of alternating between parallel sort and our
O(10)-liner node update kernel. In particular, our algorithm never
needs to deal with issuing per-sub-tree launches, with adapting launch
dimensions based on sub-tree sizes, with special-case code for
sub-trees that are ``small enough'' to merit special code paths, etc.
Though we absolutely \emph{have} looked into such codes---and have
often managed to outperform the above algorithm---the ability of
\emph{not} having to deal with more complicated method is a major
advantage of our algorithm: Almost the entire complexity and cost of
our algorithm lies in effective and stable parallel sorting, for which
many good solutions already exist.

The simplicity of our algorithm also makes it excellently suited to
templating over different data types: while our sample codes focus on
\code{float3} or \code{float4} data types we also have a variant that
is templated over arbitrary CUDA structs that can contain
arbitrary-dimensional points over arbitrary scalar types, including
arbitrary other per-point ``payload'' (such as, for example, photon
power for photon mapping~\cite{jensen2001realistic}).

This simplicity also makes it trivial to port this algorithm to other
languages, or hardware platforms. For example, the same algorithm
trivially also ports to the CPU, where it can be run with any serial
or parallel sort algorithm, and with the kernel update using, for
example, TBB for parallelism.

\section{Conclusion}

In this paper, we have described an algorithm that can build a
left-balanced k-d tree over any number of $N$ $k$-dimensional data
points in $O(N \log^2 N)$ serial complexity (and $O(\frac{N}{K}\log^2
N)$ parallel complexity, for $K$ processors). Our algorithm consists
of four different but inter-operating ideas: (1), tagging each
point with an int that stores that node's level-$l$ ancestor (and
refining this value in every iteration); (2), careful observations
regarding how to easily and efficiently compute sub-tree size and
pivot position for any given node; (3), a carefully crafted sort order
that makes these computations possible; and (4), making significant use
of parallel sort for re-arranging nodes.

We observe that the combination of tagging nodes by sub-tree ID and
sorting to produce sequential arrays of same sub-tree ID allows our
algorithm to completely avoid the need to \emph{ex}plicitly track
which sub-trees still need building, which lists of nodes these
contain, etc. This in particular allows our algorithm to always, in
each step, operate on \emph{all} elements of the array in parallel,
without the need to explicitly issue possibly many launches for
different ranges of array elements---which makes the final algorithm
outstandingly simple compared to existing list-based methods. This
ability to build the tree in a predictable $\log N$ steps that each
operate on all elements of the array make our algorithm a good match
for highly parallel architectures such as GPUs.

We make no claims that this algorithm was necessarily the fastest
possible construction method for balanced k-d trees; however, our
experiments show it to nevertheless be quite fast, allowing to build
10 million points in well under a second. This speed, combined with
how easy it is to implement this algorithm on a given architecture,
makes us believe this algorithm to be a useful tool for anybody that
needs to build of left-balanced k-d trees in a parallel and/or GPU
friendly manner.

\bibliographystyle{alpha}
\relsize{-1}{
  \bibliography{references} 

\begin{thebibliography}{MCS15}

\bibitem[BH12]{thrust}
Nathan Bell and Jared Hoberock.
\newblock {Thrust: A productivity-oriented library for CUDA}.
\newblock In {\em GPU computing gems Jade edition}, pages 359--371. Elsevier,
  2012.

\bibitem[Jen01]{jensen2001realistic}
Henrik~Wann Jensen.
\newblock {\em {Realistic image synthesis using photon mapping}}, volume 364.
\newblock Ak Peters Natick, 2001.

\bibitem[MCS15]{bitonic-sort}
Qi~Mu, Liqing Cui, and Yufei Song.
\newblock The implementation and optimization of bitonic sort algorithm based
  on {CUDA}.
\newblock {\em CoRR}, abs/1506.01446, 2015.

\bibitem[Wal22]{github-repo}
Ingo Wald.
\newblock \code{cudaKDTree}: {Sample} code for {GPU} {Construction} and
  {Stack-free} {Traversal} of {kd-Trees}, 2022.
\newblock \url{https://github.com/ingowald/cudaKDTree}.

\bibitem[Wik]{wiki-kdtree}
Wikipedia.
\newblock k-d tree.
\newblock \url{https://en.wikipedia.org/wiki/K-d_tree}, last visited Oct 23,
  2022.

\end{thebibliography}
  }

\appendix
\section{End Notes}

\begin{itemize}
    \item At the original time of writing i used \code{thrust} for the sorting, which is reflected in both this text and accompanying sample code. As of today, I would strongly suggest to use the CUDA "cub" library's \code{cub::sort} instead.
\end{itemize}

\section{Revision History}

\subsection*{Revision 3 (also April 4, 2023; now ArXiv version v4)}

\begin{itemize}
\item 
As found by same reader as previous one, in Section~\ref{sec:sb}, code snippet \emph{wrongly} stated 
\begin{lstlisting}
  ...
  /* num levels */ L = 31-clz(N+2)
\end{lstlisting}
should have been
\begin{lstlisting}
  ...
  /* num levels */ L = 32-clz(N)
\end{lstlisting}
\item changed all in-text code from \code{sc} fonts to \code{tt}, to better match code listings.
\end{itemize}

\subsection*{Revision 2 (April 4, 2023; ArXiv version v3)}

\begin{itemize}
\item 
In Section~\ref{sec:ss}, code snippet \emph{wrongly} stated 
\begin{lstlisting}
    numNodes_s_on_lowest_level = min(max(0,N-fllc_s), 1<<(N-l-1))
\end{lstlisting}
... which should have been
\begin{lstlisting}
    numNodes_s_on_lowest_level = min(max(0,N-fllc_s), 1<<(L-l-1))
\end{lstlisting}
(reference code and text were correct). A big thanks to the attentive and meticulous reader that reported this!
\item Same \code{N}-vs-\code{L} copy-n-paste error as previous one also in code snippet for
\begin{lstlisting}
    fllc_s = ...
    ss_s = ...
\end{lstlisting}
\item Abstract used regular-text "log" in "$log N$" function instead of  math-mode "$\log N$". 
\end{itemize}

\subsection*{Revision 1 (ArXiv version v2)}
  \begin{itemize}
  \item Updated title (from ``A GPU-friendly, Parallel, and (Almost-)In-Place Algorithm for Building Left-Balanced kd-Trees'' to ``GPU-friendly, Parallel, and (Almost-)In-Place Construction of
  Left-Balanced k-d Trees'')
  \item Fixed missing URLs in references.
  \end{itemize}

\subsection*{Original Version (ArXiv version v1)}
  
Originally uploaded to ArXiv Oct 31,
2022. \url{https://arxiv.org/abs/2211.00120}

\end{document}